\documentstyle[preprint,aps,prb]{revtex}
\begin{document}

\draft

\preprint{IUCM99-006}

\title{Broken Symmetry Ground States in $\nu=2$ Bilayer Quantum Hall
Systems}

\author{ A.H. MacDonald$^{1}$, R. Rajaraman$^{1,2}$ and
 T. Jungwirth$^{1,3}$}

\address{$^1$Department of Physics, Indiana University, Bloomington,
         IN 47405 \\ $^2$ School of Physical Sciences, Jawaharlal Nehru
         University, New Delhi 110067, India \\ $^3$
	 Institute of Physics ASCR, Cukrovarnick\'a 10, 162 00 Praha 6,
	 Czech Republic}

\maketitle

\begin{abstract}

We report on a study of a bilayer two-dimensional
electron gas at Landau level filling factor $\nu=2$.
The system exhibits both magnetic and spontaneous interlayer
phase coherence broken symmetries.  We propose a 3-parameter
Slater determinant variational wavefunction which
describes  the ground state over the full range
of bias potential ($\Delta_V$) 
Zeeman coupling ($\Delta_z$) and interlayer tunneling
amplitude ($\Delta_t$) strengths.  Broken symmetry states occur inside a
volume in this three-dimensional phase diagram near the
$\Delta_z^2=\Delta_V^2+\Delta_t^2$ surface.  We have obtained analytic
results for the intersections of the phase diagram with the
$\Delta_t=0$, $\Delta_z=0$, and $\Delta_V=0$, planes and show that
the differential capacitance of the bilayer system is
singular at the phase boundary.

\end{abstract}

\pacs{PACS numbers: 73.40.Hm, 32.80.Rm, 42.50.Md, 03.65.Sq}

\section{Introduction}

The interaction physics of two-dimensional electrons in the quantum
Hall regime\cite{ahmintro} is enriched by the macroscopic
degeneracy of zero-width
Landau bands.  Among the consequences of electron-electron interactions
in the quantum Hall regime are broken symmetry
ground states that are especially robust at integer Landau level filling
factors.  In a single-layer two-dimensional electron gas,
the ground state at Landau level filling factor $\nu =1$
is a strong ferromagnet\cite{qhferro} with total spin quantum number $S=N/2$
where $N$ is the number of electrons.
When mixing between different orbital Landau levels is neglected,
the exact ground state is known and consists of a fully occupied
set of lowest-Landau-level (LLL) orbitals sharing a common spinor
characterized by an arbitrary\cite{zeeman} spin orientation.  This broken symmetry
ground state is conveniently described using the lexicon of
the Hartree-Fock mean-field approximation, which happens to be
exact in this instance.  In this language, the
broken symmetry is a consequence of exchange interactions which
spontaneously spin-split the orbital Landau levels.  The
Stoner criterion for ferromagnetism in
itinerant-electron mean-field theory, $ I g(\epsilon_F) > 1$
where $I$ is the exchange integral and $g(\epsilon_F)$ is the
density-of-states at the Fermi energy, 
is always satisfied because the density-of-states
is infinite for a zero-width band.

For bilayer quantum Hall systems, described schematically in
Fig.~\ref{doublewell}, the Hartree-Fock approximation
ground state at $\nu=1$ breaks both spin-rotational invariance
and the individual-layer charge conservation symmetry\cite{dlbrokensym},
by fully occupying an orbital with an arbitrary\cite{zeeman}
spin-orientation and spontaneously developed phase coherence
between top and bottom layers even in the absence of interlayer
tunneling and Zeeman coupling.  Bilayer systems are richer still
for Landau level filling factor $\nu=2$.  In the Hartree-Fock
approximation, two orbital Landau levels are fully occupied, the
spin and bilayer state of each described by a four-component
spinor.  In realistic circumstances electron-electron interactions can
be more important than the one-particle interaction terms
corresponding to Zeeman coupling, bias potential,
or inter-layer
tunneling energy in determining the form of these
spinors, opening the possibility for broken symmetries in the ground
state.  
In previous work\cite{zhengdassarma} it has been
established that the Hartree-Fock ground state at zero bias potential
and weak Zeeman couplings is a canted antiferromagnet,
with spin ordered moments having opposing tilts
away from the Zeeman field direction in opposite quantum wells.
Theoretical predictions for the canted antiferromagnet state are
in good agreement with the findings of inelastic light scattering
experiments\cite{pellegrini}.
Although, unlike the single layer case, the Hartree-Fock
approximation does not yield the exact ground state in bilayers,
it can be improved upon systematically\cite{joglekar} and should
reliably indicate the types of broken symmetry phases which
occur.  In this paper, we use the Hartree-Fock approximation to
describe the dependence of the ground state of $\nu=2$ bilayer
quantum Hall system on a bias voltage which moves the two layers
away from density balance.  A bias potential is readily applied {\it in situ}
and, as we show, can be used to tune the system either in to or out of
the broken symmetry region of the phase diagram.
We point out that the differential capacitance of the
bilayer system, which is relatively easy to measure
experimentally, will show singular behavior at the phase
boundaries.  In Section II we introduce a four-component spinor representation
of bilayer quantum Hall ferromagnets at 
$\nu=2$.
We propose a
general 3-parameter Slater determinant variational wavefunction
which
accurately describes the ground state over the entire phase space.
We present an explicit expression for the ground state energy functional
and the four-dimensional Hartree-Fock
Hamiltonian whose eigenvalues can be used to estimate the gap
for charged excitations.  Sections III,
IV and V report analytical results for the cases of
zero Zeeman coupling, zero bias potential, and zero interlayer
tunneling, respectively.  Section III and V also include analyses
of the singularity which occurs 
in the bias-potential-dependent internal capacitance
at the order-disorder phase transition.
In Section VI we report numerical results for the
dependence of several quantities which are open to experimental
study on model parameters.  Finally in Section VII, we conclude
with a discussion of the dimension of the order parameter space. 

\section{Hartree-Fock Variational Wavefunctions}

Bilayer models for quantum well electron systems apply
whenever the growth-direction degree of freedom of the electron is adequately
described by specifying the well in which an electron resides, or equivalently
whenever the two lowest electric subbands are well separated from higher
subbands.  We adopt the bilayer model throughout this article.
The single-particle Hilbert space is then a direct product
of the two-dimensional orbital continuum space, the two-levels
associated with the electronic spin degree of freedom, and a
second two-level system which constitutes the growth-direction layer degree of
freedom. It is convenient and has become a standard notation to describe
orbital two-level systems using a pseudospin language.
In this paper we will treat the layer degree of freedom as a 
pseudospin and use a convention  in which an electron in the
top layer is said to have pseudospin up while an electron in the
bottom layer is said to have pseudospin down.  The
magnetic field is assumed to be sufficiently strong that only 2D electron
orbitals in the LLL can be occupied.  We
further assume that translational invariance is not broken in
the ground state.  We seek the single-Slater-determinant many-particle wavefunction
which minimizes the expectation
value of the Hamiltonian.  It follows from the above that it
has the form
\begin{equation}
|\Psi[z]\rangle = \prod_{i,X} \big(\sum_{k=1,4} z_k^{i}
c_{k,X}^{\dagger}\big) |0\rangle. \label{HFstate}
\label{varwf}
\end{equation}
Here $c_{k,X}^{\dagger}$ is a creation operator for a Landau gauge
LLL orbital.  The index $k$ specifies the spin and
pseudospin state; in our convention $k=1$ for an up-spin electron
in the top layer, $k=2$ for a down-spin electron in the top layer,
$k=3$ for an up-spin electron in the bottom layer and $k=4$ for a
down-spin electron in the bottom layer.  Translational invariance
requires that the coefficients $z_k^{i}$ are independent of the label
$X$ which specifies the guiding center state within LLL.
At $\nu=2$, the index $i=1,2$ and
the ground state consists of two filled Landau levels
specified by  four-component spinors $z^{1}$ and $z^{2}$, respectively.  
This is an enormous simplification
but, at first sight, still leaves us 
with a sixteen-dimensional space specified by eight complex
numbers in which to search for a minimum.  The considerations we use to
reduce this dimension are dependent in part on the form of the
Hamiltonian which we now specify.

The model we examine allows for external bias potential,
tunneling between the quantum
wells, and Zeeman coupling, in
addition to electron-electron interactions.  We write the
single-particle Hamiltonian in the form
\begin{equation}
h^0 = - (\Delta_V/2) \tau^{z} - (\Delta_t/2) \tau^{x} - (\Delta_z/2) \sigma^{x}.
\label{onebody}
\end{equation}
We have introduced the notation $\tau^{\alpha}$ for the Pauli
spin matrices acting on the pseudospin degree of freedom and
$\sigma^{\alpha}$ for the those acting on the spin degree of
freedom.  In the current context these are four-dimensional
matrices whose explicit form is easily constructed from our
convention for the $k$ indices specified above.  The parameter
$\Delta_V$ represents the electric potential drop between the two
quantum wells, due to electric field created by charges external
to the bilayer system, $\Delta_t/2$ is the amplitude for an electron
tunneling between top and bottom quantum wells which
can\cite{swierkowski} be measured by studying weak-field
Schubnikov-deHaas oscillations, and $\Delta_z$ is the single-particle
spin-splitting due to Zeeman coupling.  The single-particle
terms yield four macroscopically degenerate Landau levels whose
energies are
\begin{equation}
E^0_{1,2,3,4} =  \pm\frac{1}{2} 
\left[ \sqrt{\Delta_t^2+\Delta_V^2}\pm \Delta_z \right]
\label{epsilon0}
\end{equation}
Broken symmetry states for $\nu=2$ occur only when the relative and absolute
strengths of the three single-particle terms is such that the
gap between the two lower energy eigenvalues and the two higher
energy eigenvalues is smaller than or comparable to
electron-electron interaction strengths.

The many-particle
Hamiltonian whose expectation value we minimize is
\begin{eqnarray}
\hat{H} &=& \sum_{k',k,X} c_{k',X}^{\dagger} h^0_{k',k} c_{k,X}
+ \frac{1}{2} \sum_{k1,k2} \sum_{X_1,X_2,X_1',X_2'}
c_{k1,X1'}^{\dagger} c_{k2,X2'}^{\dagger} c_{k1,X1} c_{k2,X2}
\big[ \langle X_1^{'},X_2^{'}|V_{+}|X_1,X_2\rangle \nonumber \\
 &+& \tau^{z}_{k1,k1}
\tau^{z}_{k2,k2}\langle X_1^{'},X_2^{'}|V_{-}|X_1,X_2\rangle \big].
\label{manyparticle}
\end{eqnarray}
Here $V_{\pm} = (V_S \pm V_D)/2$ and $V_S$ and $V_D$ are the 2D
interactions between electrons in the same layer and in different
layers.  The qualitative physics of bilayer quantum Hall systems at $\nu=2$ is
controlled by the property that $V_D$ is weaker than $V_S$.
Note that this model explicitly drops the, normally small, terms
in which interactions scatter electrons between layers.  The model
also assumes that the two quantum wells in the bilayer system are
identical, usually true of in experimental systems.  This assumption
can easily be relaxed and does not change any qualitative physics.

The properties of orbital states\cite{harfok}
within a Landau level can be used to 
derive a general expression for the variational energy of the 
Slater determinant trial wavefunctions:
\begin{eqnarray}
\epsilon &=& -\frac{1}{2} \sum_{i=1}^{2} \big[
\Delta_V \ \langle z^{i} | \tau^{z} | z^{i} \rangle + 
\Delta_t \langle z^{i} | \tau^{x} | z^{i} \rangle
+ \Delta_z \ \langle z^{i} | \sigma^{x} | z^{i} \rangle  \nonumber \\
&+&   F_{+} - \sum_{j=1}^{2} \left(H
\langle z^{i} | \tau^{z} | z^{i} \rangle 
\langle z^{j} | \tau^{z} | z^{j} \rangle
-F_{-} \langle z^{i} | \tau^{z} | z^{j} \rangle 
\langle z^{j} | \tau^{z} | z^{i}
\rangle \right)\big]
\label{eng}
\end{eqnarray}
Here $\epsilon$ is energy per Landau level orbital.  (The number of 
orbitals in a Landau level is $N_{\phi} = A B / \Phi_0$ where 
$A$ is the cross-sectional area and $\Phi_0$ is the electron 
magnetic flux quantum.)  In Eq.[~\ref{eng}] we have adopted Dirac
notation for the two four-component
spinors and introduced the three interaction energy parameters which 
appear in the Hartree-Fock theory of bilayer quantum Hall systems:
\begin{equation}
H = (2 \pi \ell^2)^{-1} V_{-}(\vec q = 0),
\label{hartree}
\end{equation}
and
\begin{equation}
F_{\pm}  \ = \ \int \frac{d^2 \vec q}{(2 \pi)^2}
\exp ( - q^2 \ell^2/2) V_{\pm} (\vec q).
\label{fock}
\end{equation}
The parameter $H$ characterizes the electrostatic energy
associated with charge transfer between the wells, while $F_{+}$
and $F_{-}$ are exchange energies associated with the sum and
difference of intralayer and interlayer interactions. Note that
$F_{+}$ contributes
only a constant to the energy so that the ground state depends only on
the three single-particle parameters and on $H$ and $F_{-}$.  In the
limit of infinitely narrow quantum wells and for the LLL,
$H =(e^2/\varepsilon\ell) d / 2 \ell$ where $\varepsilon$ is the
dielectric constant and $\ell = (\hbar c /e B)^{1/2}$
is the magnetic length. The
interaction parameters depend in general on details of the bilayer system
geometry and on the index of the orbital Landau level at the Fermi
energy.  The values of these parameters are easily evaluated once
the geometry has been specified.  We assume in what
follows\cite{remark} that $H > F_{-}$,
an inequality which is always satisfied when the
interacting 2D electron layers are identical.  

Extrema of the Hartree-Fock variational energy functional satisfy
the Hartree-Fock single particle equations.  For the present
problem, the spinors we seek are the two lowest energy eigenstates
of the four-dimensional matrix
\begin{equation}
h^{HF} = h^0 -F_{+}\rho + H \tau^{z} [{\rm Tr}(\rho \tau^{z})]
 - F_{-} \tau^{z} \rho \tau^{z}
\label{hhf}
\end{equation}
where $\rho$ is the four-dimensional density matrix which must
be constructed self-consistently:
\begin{equation}
\rho_{k',k} = \sum_{i=1}^{2} \langle b^{k'} | z^i \rangle 
\langle z^i | b^k
\rangle
\label{densmat}
\end{equation}
Here $b^k$ denote the four four-component basis spinors 
($b^1$=(1,0,0,0), $b^2$=(0,1,0,0), $b^3$=(0,0,1,0), $b^4$=(0,0,0,1)) 
corresponding to label-$k$ spin and pseudospin states as introduced 
in Eg.~(\ref{varwf}).
The Hartree-Fock variational state can be determined by finding
solutions of the Hartree-Fock equations or directly by minimizing
the energy functional (\ref{eng}) with parameterized spinors.   The former
approach is usually more convenient when numerical solutions are
required and has the advantage of yielding an estimate of the quasiparticle
excitation
gap, which is important in interpreting transport
experiments\cite{ezawarecent}.  
The latter approach is often more convenient when the calculation can
be done analytically.
In this paper we will try to proceed as far as possible analytically
before resorting to numerical calculations  to  illustrate and extend
the analytical results.

We start from a variational
{\em ansatz} \, for the two four-component
spinors $z^{1}$ and $z^{2}$  appearing
in the $\nu=2$ translationally invariant trial wavefunction (Eq.[~\ref{varwf}]).
(A more complete discussion of the most general translationally
invariant $\nu = 2$ wavefunction and of the number of parameters it can carry
will be given in the concluding Section VII.)  
We have checked numerically to ensure
that our ansatz, characterized by three variational
parameters, is sufficiently general to capture the minimum energy state for any
values of the parameters which appears in the one-body 
Hamiltonian (\ref{onebody}), namely, the bias potential,
the Zeeman coupling and the tunneling amplitude.
Our wavefunctions are specified by the following two four-component
spinors:
\begin{eqnarray}
 z^{1} \  &=& (C c_{\uparrow}, S s_{\downarrow}, C s_{\uparrow}, -S
c_{\downarrow}) \nonumber \\
 z^{2} \ &=& (-S s_{\uparrow}, C c_{\downarrow}, S c_{\uparrow},
Cs_{\downarrow}) \label{generalwf}
\end{eqnarray}
Here $C = \cos (\theta/2)$, $S=\sin(\theta/2)$,
$c_{\sigma}=\cos(\chi_{\sigma}/2)$, and
$s_{\sigma}=\sin(\chi_{\sigma}/2)$ with $\sigma \ = \ \uparrow \ , \downarrow$.
The variational energy of these wavefunctions is
\begin{eqnarray}
\epsilon &=&
-\Delta_V  \cos(\theta) \cos(\chi_{+}) \cos(\chi_{-}) 
-\Delta_t \cos(\theta) \sin(\chi_{+}) \cos(\chi_{-})
-\Delta_z \sin(\theta) \sin(\chi_{-}) 
\nonumber \\
&-&F_{+}
-F_{-} [\sin^2(\theta) \sin^2(\chi_{-}) 
+ \cos^2(\theta) (\cos^2(\chi_{+}) \cos^2(\chi_{-})+ \sin^2(\chi_{+}) \sin^2(\chi_{-}))]
 \nonumber \\
 &+& 2 H \cos^2(\theta) \cos^2(\chi_{+}) \cos^2(\chi_{-})
\label{generalenergy}
\end{eqnarray}
Here $\chi_{\pm} \equiv (\chi_{\downarrow} \pm \chi_{\uparrow})/2$.
We have not been able to minimize this energy expression analytically 
for the case when all three parameters $\Delta_V, \ 
\Delta_t$  and $\Delta_z$ are non-zero.
In the following sections we present analytic results for the cases 
in which one of the 
three parameters is set to zero.  

The density matrix corresponding to our variational wavefunction is 
\begin{eqnarray}
2 \rho &=& {\protect\bf 1} + \cos(\theta) \cos(\chi_{+}) 
\cos(\chi_{-}) \tau^{z} 
 + \sin(\theta) \sin(\chi_{-}) \sigma^{x}   \nonumber \\
 &+& \cos(\theta) \sin(\chi_{+}) \cos(\chi_{-}) \tau^{x}   
 + \sin(\theta) (\cos(\chi_{-}) - \cos(\chi_{+})) \sigma^{x} \tau^{x} 
 \nonumber \\
 &-& \cos(\theta) \cos(\chi_{+}) \sin(\chi_{-}) \sigma^{z} \tau^{x} 
 + \cos(\theta) \sin(\chi_{+}) \sin(\chi_{-}) \sigma^{z} \tau^{z} 
\label{dmansatz}
\end{eqnarray} 
Expectation values of spin and pseudospin operators and their products can
be read off from Eq.~(\ref{dmansatz}) by using the familiar properties of 
Pauli matrices.  For example,  
the difference between the $\hat z$ direction projection of the
spin in the top and bottom layers, $ \langle \tau^{z} \sigma^{z} \rangle 
= Tr(\rho \tau^{z} \sigma^{z} )$  which was identified as the order
parameter in the work of Zheng {\it et al.}\cite{zhengdassarma},
is given by 
\begin{equation}
O_{zz} = 2 \cos(\theta) \sin(\chi_{+}) \sin(\chi_{-}).
\label{orderparam}
\end{equation}
This order parameter is like that of a canted antiferromagnet; 
indeed this
is the rubric used by Zheng {\it et al.} to describe the broken symmetry
state.  We will show that 
\begin{equation} 
O_{xz} \equiv  - \langle \tau^{x} \sigma^{z} \rangle = 2
\cos(\theta) \cos(\chi_{+}) \sin(\chi_{-}) 
\label{neworderparam}
\end{equation}
can  also be non-zero.  We discuss the meaning of this new order parameter,
which  is non-local in the growth-direction spatial coordinate,
at greater length later.  
Finite expectation values for either operator require broken symmetry
since the Hamiltonian is invariant under rotations about the $\hat x$ 
axis in spin space.

Given the density matrix~(\ref{dmansatz}), 
the Hartree-Fock single particle Hamiltonian is 
readily evaluated:
\begin{eqnarray}
2 h^{HF} &=& -F_S + (4H - F_S - \Delta_V) 
\cos(\theta) \cos(\chi_{+}) \cos(\chi_{-}) \tau^{z} 
 - (F_S + \Delta_z) \sin(\theta) \sin(\chi_{-}) \sigma^{x} \nonumber \\ 
 &-& (F_D + \Delta_t) \cos(\theta) \sin(\chi_{+}) \cos(\chi_{-}) \tau^{x}
 + F_D \cos(\theta) \cos(\chi_{+}) \sin(\chi_{-}) \sigma^{z} \tau^{x} 
 \nonumber \\
 &-& F_D \sin(\theta) (\cos(\chi_{+}) - \cos(\chi_{-})) \sigma^{x} \tau^{x} 
 - F_S \cos(\theta) \sin(\chi_{+}) \sin(\chi_{-}) \sigma^{z} \tau^{z} \; ,
\label{harfokexp}
\end{eqnarray}
where $F_S=F_++F_-$ and $F_D=F_+-F_-$.
When the variational parameters are at an extremum of the Hartree-Fock
energy functional~(\ref{generalenergy}), 
$z^{1}$ and $z^{2}$ are eigenvectors of  
matrix~(\ref{harfokexp}) and
if the extremum is the minimum 
these eigenvectors will correspond to the two 
lowest energy eigenvalues, $E^{HF}_1$ and $E^{HF}_2$.  
The Hartree-Fock approximation
for the charged excitation energy gap, measured by quantum Hall 
transport experiments, is $E_{gap}=E^{HF}_{3}-E^{HF}_{2}$. 

\section{The Vanishing Zeeman Coupling Limit}

At  $\Delta_z = 0$, we expect  the ground state
spin-magnetization directions in the two quantum wells to be in opposition.  
The collinear spin configuration corresponds to taking the parameter
$\theta$ to be zero in the ansatz (\ref{generalwf}). This
 reduces the number of free
parameters in the variational wavefunction to two, namely,
 the polar angles
for the up-spin and down-spin pseudospin orientations.
With $\theta = 0$, the spinors (\ref{generalwf}) reduce to
$z^{1}=(\cos(\chi^{\uparrow}/2),0,\sin(\chi^{\uparrow}/2),0)$
and $z^{2}=(0,\cos(\chi^{\downarrow}/2),0,\sin(\chi^{\downarrow}/2))$,
and the energy per orbital state (\ref{generalenergy}) reads

\begin{eqnarray}
\epsilon &=&  - \Delta_V \cos(\chi_{+}) \cos(\chi_{-})
+ 2 H \cos^2(\chi_{+}) \cos^2(\chi_{-}) \nonumber  \\ 
&-&  \Delta_t \sin(\chi_{+}) \cos(\chi_{-}) 
- F_{+}  \ - \ F_{-} (\cos^{2}(\chi_{+}) \cos^{2}(\chi_{-})
+ \sin^{2}(\chi_{+}) \sin^{2}(\chi_{-})) \; .
\label{energy}
\end{eqnarray}

The first two terms on the right hand side of this equation represent the
electrostatic energy which is minimized when the `flat band'
condition, $\cos(\chi^{+}) \cos(\chi^{-}) =
\Delta_V/4H$, is satisfied.   If these were the only terms in the energy
expression, the charge transfer between layers would perfectly screen
the external interlayer electric field parameterized here by $\Delta_V$.
Since we have not included Zeeman coupling, any difference between
the pseudospin angles of up-spins and down-spins, {\i.e.} $\chi_{-} 
\ne 0$, represents a broken symmetry.
The exchange energy is maximized in magnitude by localizing both
up-spins and down-spins in one of the layers; at $\Delta_V=0$ 
this
objective can be made consistent with minimizing the electrostatic
energy by choosing opposite layers for the two spins, i.e.,
$\chi_{+} =\pi/2$.  
Minimizing with respect to $\chi_{-}$ 
then gives $\cos(\chi_{-}) =  \Delta_t/2F_{-}$
for $\Delta_t < 2F_{-}$.   Otherwise $\cos(\chi_{-}) = 1$
and there is no broken symmetry. The spin-up and spin-down pseudospins 
for $\Delta_z=0$, $\Delta_V=0$  are
schematically illustrated in Fig.~\ref{dtdv}. Here and in all figures
below, we plot results for layer separation $d=1$ in units 
of the magnetic length $\ell$
yielding the following interaction strengths: $H=0.5$, $F_+=0.9545$, and
$F_-=0.2988$, where energy is measured in units of $e^2/\varepsilon\ell$.  
The mean pseudospin
orientation is in the $\hat{x}$ direction which corresponds to equal 
distribution of charge between the layers. For $\Delta_t > 2F_-$
the two pseudospins are parallel ($\chi_- = 0$) and the spin polarization
in each  well is zero, as also indicated in Fig.~\ref{dtdv}. For 
$\Delta_t< 2F_-$, $z^1$ and $z^2$ have pseudospin orientations 
tilted from the $\hat{x}$ direction
upwards and downwards, respectively, $\chi_-\ne 0$ and, hence, 
the order parameter
$O_{zz}$ is non-zero. As the tilt angle increases with decreasing $\Delta_t$,
opposite spin polarizations builds up in the two layers reaching
a maximum as $\Delta_t\rightarrow 0$.  Then the two pseudospins are in
opposition and aligned in the $\hat{z}$ direction. The non-zero
value of $O_{zz}$ in the ordered region reflects the broken SU(2)
symmetry (full rotational symmetry) of spin space. Note that 
the order parameter $O_{xz}$ is zero along the $\Delta_z=0$, $\Delta_V =0$
line since $\chi_+ = \pi/2$.

We now turn our attention to $\Delta_z=0$, $\Delta_t = 0$
line in the phase diagram. In this case the energy is minimized when 
one of the spinors, e.g.  $z^1$, is localized in the top well, i.e., 
$\chi_{\uparrow}=0$ so that  
$\chi_-=\chi_+=\chi_{\downarrow}$.
Minimizing the energy
functional (\ref{energy}) with respect
to $\chi_-$ gives
\begin{equation}
\cos^2(\chi_-)=\frac{\Delta_V-2F_-}{4(H-F_-)}
\label{deleq0}
\end{equation}
for $2F_- < \Delta_V < 4H - 2F_-$, $\cos^2(\chi_-)=0$ for
$\Delta_V< 2F_-$ and $\cos^2(\chi_-)=1$ for 
$\Delta_V > 4H-2F_-$. 
For bias potential $\Delta_V < 2 F_{-}$ 
the up and down spin orbitals are localized in opposite wells and
no charge transfer is produced, as illustrated in Fig.~\ref{dtdv}.  
For $\Delta_V > 4H -
2F_{-}$, the up and down spin orbitals are
localized in the top well and the charge transfer is complete.
At intermediate values of $\Delta_V$, Eq.~(\ref{deleq0}) applies.  
In the interval
$2F_{-} < \Delta_V < 4H - 2F_{-}$, both $O_{zz}\ne 0$ and $O_{xz}\ne 0$. 
The non-zero value of $O_{zz}$ reflects broken $SU(2)$ spin-symmetry
in the ground state while the non-zero value of $O_{xz}$ at 
$\Delta_t = 0$ reflects a $U(1)$ broken pseudospin symmetry.
At $\Delta_t = 0$ the Hamiltonian is invariant under rotations about the
$\hat z$ axis in pseudospin space; the number of particles in each layer
is a good quantum number. 
This symmetry is broken since the orbital states with variational
spinor $z^2$ possess 
interlayer phase coherence (see Fig.~\ref{dtdv}). 
As in the $\nu=1$ case, spontaneous interlayer phase coherence
will lead to a variety of interesting
effects\cite{doublelayerrefs} which we do not pursue further here.
We return to this limit again in Section V.

We have been
unable to obtain general analytic expression for the minimum energy state
for both $\Delta_V$ and $\Delta_t$ non-zero.  
Instead we define $C = \cos(\chi_{+}) \cos(\chi_{-})$, $S=
\sin(\chi_{+}) \sin(\chi_{-})$ and expand around the disordered state for
which $S=0$.  We consider $S$ as an order parameter and perform a 
Ginzburg-Landau analysis.  
\begin{equation}
\epsilon = \epsilon_0(C) + \epsilon_2(C) S^2 
+ \frac{1}{2} \epsilon_4(C) S^4  + \ldots.
\label{gleq}
\end{equation}
Comparing with Eq.~(\ref{energy}), we obtain the following results for the
coefficients of this Ginzburg Landau expansion:
\begin{eqnarray}
\epsilon_0(C) &=& -\Delta_V C + 2 H C^2 - 
\Delta_t \sqrt{(1-C^2)} -F_{+} - F_{-}C^2
\nonumber \\
\epsilon_2(C) &=& \frac{\Delta_t}{2 (1-C^2)^{3/2}} - F_{-} \nonumber \\
\epsilon_4(C) &=& \frac{\Delta_t}{4} \frac{1 + 4 C^2}{(1-C^2)^{7/2}}.
\label{eqglexp}
\end{eqnarray}
In the normal state, $S$ is zero and $C$ is 
determined my minimizing $\epsilon_0(C)$, implying the following
relationship between $\Delta_V$ and $C$:
\begin{equation}
\Delta_V = (4H-2F_{-})C +\Delta_t \frac{C}{(1-C^2)^{1/2}}
\label{vnorm}
\end{equation}
The normal state becomes unstable when $\epsilon_2(C)$ becomes negative;
on the phase boundary, $C$ equals
\begin{equation}
C^{*} = (1-(\Delta_t/2F_{-})^{2/3})^{1/2}.
\label{cstar}
\end{equation}
As the bias potential $\Delta_V$ increases, $\epsilon_2(C)$ increases,
crossing zero at the critical bias potential
\begin{equation}
\Delta_V^{*} = \big[ 4 H - 2 F_{-} + \Delta_t^{2/3} (2 F_{-})^{1/3}\big]
(1 - (\Delta_t/2F_{-})^{2/3})^{1/2}.
\label{vstar}
\end{equation}
Eq.~(\ref{vstar}) describes the projection of the mean-field phase boundary
onto the $\Delta_z =0$ plane, illustrated in   
Fig.~\ref{dtdv}.  

Among the observables which are singular at this $T=0$ phase transition, the
most experimentally accessible\cite{capexp} is the internal differential
capacitance of the bilayer system:
\begin{equation}
C_{int} \equiv \frac{d \sum_{i=1}^{2} \langle z^i | \tau^{z}|
z^i\rangle}{d\Delta_V}= 2 \frac{dC}{d\Delta_V}
\label{capdef}
\end{equation}
In the normal state, an equation for $C_{int}$ can be obtained by
differentiating both sides of Eq.~(\ref{vnorm}) with respect to
$\Delta_V$:
\begin{equation}
C_{int}^{norm} = \frac{2}{4H - 2 F_{-} + \Delta_t (1-C^2)^{-3/2}}
\end{equation}
Only the first term in the denominator appears in the
electrostatic approximation for this capacitance.  A similar
expression, valid near the phase boundary, may be derived for the
ordered state by minimizing the Ginzburg-Landau energy expression
with respect to $S$.  We find that
\begin{equation}
\epsilon \approx \epsilon_0(C) - \Lambda (C^{*}-C)^2.
\label{condeng}
\end{equation}
where
\begin{equation}
\Lambda = \frac{ 18 F_{-} (1 - (\Delta_t/2F_{-})^{2/3})}{5 - 4
(\Delta_t/2F_{-})^{2/3}}.
\label{cinvjump}
\end{equation}
The additional condensation contribution to the energy alters the
inverse capacitance. Differentiating Eq.~(\ref{condeng}) first
with respect to $C$ and then with respect to $\Delta_V$ we find that
as the phase boundary is approached from the ordered state  
side
\begin{equation}
C_{int}^{order} = \frac{2}{4H - 2 F_{-} + \Delta_t (1-C^2)^{-3/2}-
\Lambda}\; .
\end{equation}
In mean-field-theory the capacitance has a jump discontinuity at
the phase boundary.  This observable is the analog, for this
bias tuned quantum phase transition, of the specific heat at a temperature
tuned phase transition.  The differential capacitance is 
singular at all points in the two-dimensional boundary 
of the ordered state region in the three-dimensional phase space. 
This singular behavior should be accessible to experiment and 
may be interesting to study, especially since the nature of 
the broken symmetry in the ordered state crosses over from SU(2) to 
U(1) on going away from the $\Delta_z = 0$ plane.    
The mean-file critical exponent will
be replaced by the specific heat exponent of a {\it three-dimensional}
classical system, presumably that of the 3D Heisenberg model at 
$\Delta_z =0$ and that of the 3D XY model at $\Delta_z \ne 0$.
In Section V we give an analytical expression for the capacitance jump
for finite Zeeman coupling in the limit of vanishing interlayer tunneling.

\section{The Vanishing Bias Potential Limit}

The Hartree-Fock equations for the zero bias potential limit have been solved
previously by Das Sarma {\it et al.}\cite{zhengdassarma} in their
pioneering work on $\nu=2$ broken symmetry ground states.  In this
section we report analytic solutions of the Hartree-Fock equations
which provide some additional insight.
When $\Delta_V=0$, we expect the energy
minimum to occur for equal charge density in the two layers,
i.e., at $\langle \tau^{z} \rangle = 2 \cos(\theta) \cos(\chi_{+})
\cos(\chi_{-}) = 0$.   It is clear that the energy is minimized by 
$\chi_{+} = \pi/2$, since this choice does not frustrate any of the 
other energy terms.  
The Hartree energy then vanishes and the total
energy expression (\ref{generalenergy}) simplifies to
\begin{equation}
\epsilon =
-\Delta_t \cos(\theta) \cos(\chi_{-})
-\Delta_z \sin(\theta) \sin(\chi_{-})
-F_{+} - F_{-} \sin^2(\chi_{-}).
\label{veq0energy}
\end{equation}
Minimizing with respect to $\theta$ we find that at the
extremum
\begin{equation}
\Delta_t \cos (\chi_{-}) \sin (\theta) = \Delta_z \cos(\theta) \sin(\chi_{-}).
\label{ddthetaeq0}
\end{equation}
Minimizing with respect to $\chi_-$ and using Eq.~(\ref{ddthetaeq0})
we find that
\begin{equation}
(\Delta_t - \Delta_z^2/\Delta_t) \cos (\theta) = 2 F_{-} \cos(\chi_{-}).
\label{ddchieq0}
\end{equation}
These two equations can be combined to find explicit equations
for $\cos(\theta)$ and $\sin(\chi_{-})$:
\begin{eqnarray}
\cos^2(\theta) &=& \frac{ \Delta_t^2 
[ (\Delta_t^2-\Delta_z^2)^2-(2 F_{-} \Delta_z)^2]}
{(\Delta_t^2-\Delta_z^2)^3} \nonumber \\
\sin^2(\chi_{-}) &=& \frac{4 \Delta_t^2 F_{-}^2 - (\Delta_t^2-\Delta_z^2)^2}
{ 4 F_{-}^2 (\Delta_t^2 - \Delta_z^2)}.
\label{extremum}
\end{eqnarray}

Broken symmetry states occur when both of these equations have
solutions.  For $\Delta_t$ larger than $2 F_{-}$ and
$\Delta_z$ smaller than
\begin{equation}
\Delta_z^{min} = \sqrt{\Delta_t (\Delta_t - 2F_{-})}
\label{hmin}
\end{equation}
the expression for $\sin^2(\chi_{-})$ is negative.  The minimum energy
then occurs for $\sin(\chi_{-})=0$ and
$\theta =0$ and the variational
spinors reduce  to the up-spin and down-spin symmetric
states, as illustrated in Fig.~\ref{dtdz}. These states are
expected to have the lowest two energies for
non-interacting electrons for $\Delta_z < \Delta_t$.  For $\Delta_z$ larger than
\begin{equation}
\Delta_z^{max} = \sqrt{F_{-}^2 + \Delta_t^2} - F_{-}
\label{hmax}
\end{equation}
the expression for $\cos^2(\theta)$ is negative.  The minimum
energy then occurs for $\theta = \pi/2$, $\chi_{\uparrow}=0$,
and $\chi_{\downarrow}=\pi$.  In this case the variational spinors
are both positive eigenvalue eigenstates of $\sigma^{x}$.
The wavefunction in this region
is the non-interacting electron state for $\Delta_z > \Delta_t$.
In the ordered region, $O_{zz}$ is non-zero and the spin U(1) symmetry 
is broken. The magnetic configuration (see Fig.~\ref{dtdz}) 
is that of a canted antiferromagnet as observed  
by Zheng {\em et al.}.\cite{zhengdassarma} Each of the two spinors $z^1$ and
$z^2$ has non-zero spin up and spin down components in the ordered 
region for $\Delta_z >0$. However, the weights of  spin down 
component of $z^1$ and spin up component of $z^2$ 
decrease with decreasing Zeeman coupling and the state
continuously develops into the one illustrated in Fig.~\ref{dtdv}
in the limit of $\Delta_z=0$.
Note that $\Delta_z^{max} \propto \Delta_t^2$
at small $\Delta_t$ so the broken symmetry state is not robust
against Zeeman coupling when $\Delta_t$ is small.  Note also that
when $\Delta_t$ is large $\Delta_z^{min} = \Delta_t - F_{-}
-F_{-}^2/2\Delta_t + \ldots$ while $\Delta_z^{max} = \Delta_t
- F_{-} + F_{-}^2/2\Delta_t + \ldots$; the broken symmetry state
interpolates between the fully spin-polarized and fully
pseudospin polarized states and
occurs only over a narrow interval of Zeeman fields surrounding
$\Delta_z = \Delta_t - F_{-}$.  The field at which the crossover occurs
is shifted from $\Delta_t$ because of the exchange energy is 
more favorable for the full spin-polarized state 
than for the fully pseudospin-polarized state.

\section{The Vanishing Interlayer Tunneling Limit} 

For $\Delta_t =0$, we expect $\chi_{+}= 0$.  Numerical calculations
presented in the next section
verify that this choice of $\chi_{+}$ gives the optimal variational
energy for vanishing interlayer tunneling.  
Given $\chi_{+} =0$, it can be seen that the energy is minimized by
$\theta = \chi_{-}$, leaving a single parameter to be determined
by minimizing $ -\Delta_V x - \Delta_z (1-x) - F_{+}
-F_{-}((1-x)^2 + x^2) + 2 H x^2$ with respect to $ x =
\cos^2(\theta)=\cos^2(\chi_{-})$.  We find that 
$\cos(\theta) = \cos(\chi_{-}) = 0$ for $\Delta_V < 2 F_{-} + \Delta_z$,
$\cos(\theta) = \cos(\chi_{-}) = 1$ for $\Delta_V > 
4 H - 2 F_{-} + \Delta_z$, and that 
\begin{equation}
\cos^2(\theta) = \frac{ \Delta_V - \Delta_z - 2F_{-}}{4(H-F_{-})}
\label{dezeq0}
\end{equation}
in the intervening interval.  The low bias state is fully spin polarized
along the Zeeman field direction and has the charge equally distributed
between the two layers, as indicated in Fig.~\ref{dvdz}. The high bias
state is spin unpolarized and
has all electrons in the top layer.  
For non-interacting electrons, the transition
between these two states is first order and occurs  at 
$\Delta_V=\Delta_z$.  
In the Hartree approximation, charge transfer occurs linearly 
with bias voltage, since the double layer system acts like a capacitor.  
In this approximation, the minority spin
Landau level in the top layer and the majority spin Landau level in the bottom
layer are degenerate throughout the interval of bias voltages where charge
transfer takes place.  We show below that an energy gap exists throughout 
this interval, and that the gap is supported by the development of a broken
symmetry state.  However, the broken symmetry is not the canted antiferromagnet
order of the $\Delta_V=0$ plane but rather something 
closely akin to the spontaneous
interlayer phase coherence which occurs at $\nu =1$ in double-layer 
systems.  The projection of the phase diagram onto the $\Delta_t =0$ plane 
is illustrated in Fig.~\ref{dvdz}.

In the ordered portion of the $\Delta_t=0$ plane, $O_{zz}=0$ and  
\begin{equation}
O_{xz} = \frac{1}{2} \big[ 1 - \big(\frac{\Delta_V - \Delta_z - 2 H}{2 (H-F_{-}}\big)^2 \big]^{1/2}.
\label{onl}
\end{equation}
In the ordered state, the Landau level orbitals with spinors
$z^{1}$ and $z^{2}$
spontaneously develop interlayer phase coherence and spin-polarization in the 
$\hat z$ direction.  (The Zeeman field is oriented in the $\hat x$ direction.)
One Landau level orbital has up and down spin projections whose pseudospin
orientations are proportional to $[\sin(\chi_{-}),0,\cos(\chi_{-})]$ and 
$[\sin(\chi_{-}),0,-\cos(\chi_{-})]$ respectively; the spin projections of
the other Landau level orbital have opposing pseudospin 
orientation and altered weights (see Fig.~\ref{dvdz}).
Spontaneous interlayer phase coherence corresponds to non-zero 
pseudospin components in the $\hat x$ direction.  The occurrence of 
this type of order is driven by interlayer interactions; the
spatial correlations
between electrons in different layers are improved when it is established.

We note that the solutions we have found for the $\Delta_t \to 0$ limit of the 
$\Delta_z =0$ plane and the $\Delta_z \to 0$ limit of the 
$\Delta_t =0$ plane differ.  Evidently, the order of limits matters 
in this instance.  It turns out that when both these terms
are absent, energy minimization fixes a value only
for the product $\cos(\theta) \cos(\chi^{+})$ and not for 
these factors individually.  In the $\Delta_t = 0$ plane, the state is 
invariant under a simultaneous rotation of all pseudospins around
the $\hat z$ axis and under a rotations of spins around the $\hat{x}$ axis.
These invariances are explicitly accounted for in limiting the number of 
parameters in our variational wavefunction. 
For $\Delta_z =0$, the state is invariant under rotation about any
spin-axis.  When both external fields are set to zero, all symmetries are 
present, leading to additional soft modes and a set of variational 
wavefunctions which have the same energy.  When these field parameters 
approach zero, $O_{zz} \gg O_{xz} $ if $\Delta_t \gg \Delta_z$, 
and $O_{xz} \gg O_{zz} $ if $\Delta_z \gg \Delta_t$.  From
the above solution for the variation parameters, it is possible to show
that for $\Delta_t = 0$, $E_{gap} = F_{D}$, a constant throughout
the ordered region of the phase diagram.
The gap arises entirely because of the interlayer interactions
which give rise to spontaneous interlayer phase coherence.

In the $\Delta_t=0$ plane, $\langle\tau^z\rangle=2\cos^2(\theta)$. Using 
Eq.~(\ref{dezeq0}) we can derive an analytical expression for the internal
capacitance in the ordered phase:
\begin{equation}
C_{int}^{order}=\frac{d\langle\tau^z\rangle}{d\Delta_V}=\frac{1}{2(H-F_-)}\; .
\label{capdt0}
\end{equation}
In the normal state, no charge transfer is produced by changing $\Delta_V$
in either  low or high bias regions of the phase diagram. The internal 
capacitance is hence zero throughout the normal state region and has a jump,
given by (\ref{capdt0}), along the $\Delta_V=\Delta_z+2F_-$ and 
$\Delta_V=\Delta_z+4H-2F_-$ lines.

\section{Numerical Results}

We supplement the analytical results discussed in the previous sections 
with results obtained from numerical solutions of the four-dimensional
Hartree-Fock equation outlined in Section II.  An  overview on the 
order-disorder physics of bilayer quantum Hall systems at $\nu=2$ is
provided in Fig.~\ref{3d}, a three-dimensional phase diagram
whose axes are the three one-body external fields, $\Delta_V$,
$\Delta_t$, and $\Delta_z$.  In this diagram, the region with
spontaneous order is enclosed by two surfaces.
In the limit of no interactions these collapse onto the single 
surface, $\Delta_V^2+\Delta_t^2 = \Delta_z^2$, where the 2nd and 3rd 
Landau levels are degenerate.  {\it Broken symmetry states can occur
at integer filling factors when interactions dominate the energetic
splitting of nearby Landau levels.}  The width of the ordered region tends 
to expand when the single-particle field parameters are small and 
all four single-particle Landau levels are close to degeneracy.
The intersection of this phase diagram
with the $\Delta_V=0$ plane recovers the results obtained originally
by Das Sarma {\it et al.} and derived analytically in Section IV.
In the ordered region within the 
$\Delta_V=0$ plane, $\sin(\chi_{+}) =1$, and  $O_{zz} \ne 0$
while $O_{xz} =0$.
When $\Delta_t \ne 0$, interlayer phase coherence is not spontaneous
so $O_{zz} \ne 0$ represents only the $\hat x$-axis
spin-rotation (U(1)) broken symmetry. 
In the ordered portion of the $\Delta_t =0$ plane and for $\Delta_z>0$,
$O_{xz} \ne 0$, but
$O_{zz}=0$.  The order here  involves spontaneous interlayer 
phase coherence and spin order; as at $\nu=1$ 
we can expect interesting physics to occur for small $\Delta_t$ 
in tilted fields in this region of the phase diagram.
The ordered state has $\hat x$-axis spin-rotation
and $\hat z$-axis pseudospin-rotation broken symmetries 
(\, U(1)$\otimes$U(1) broken symmetry)
so we expect two distinct Goldstone modes to appear. At $\Delta_z=0$ the
broken symmetry changes to SU(2)$\otimes$U(1) in the ordered region
of $\Delta_t=0$ plane.
The phase diagram intersections with the $\Delta_z=0$ and $\Delta_t=0$ planes 
are in agreement with the analytic results presented in 
Sections III and V respectively.  In the ordered region of the 
$\Delta_z=0$ plane, both $O_{zz}$ and $O_{xz}$ are non-zero and 
are  indices of the same broken symmetry.  
In this case,
the Hamiltonian is spin-rotationally invariant so we have a 
SU(2) broken symmetry.

In experimental samples, 
$\Delta_V$ and $\Delta_t$ can easily be made large but $\Delta_z$ is 
usually small.  Since it is easy to tune the bias voltage $\Delta_V$ 
over large 
ranges experimentally,  we examine below the dependence of physical observables
on this parameter with the other two parameters held fixed.

In Figs.~\ref{numerical}(a)-(c) we plot the bias dependence 
of the following quantities,
all of which are experimentally
accessible: the two order parameters $O_{zz}$ and $O_{xz}$,
the Hartree-Fock energy gap $E_{gap}$, the difference between top layer
and bottom layer charge densities $\langle \tau^z \rangle$, and 
the differential capacitance of the bilayer system,
$d \langle \tau^z \rangle / d\Delta_V$.  The bias dependence of the energy
gap has been extensively studied experimentally;\cite{ezawarecent} 
we expect the present Hartree-Fock theory to capture the 
main trends\cite{demler} in these observations although we 
know that spin-texture excitations can be important\cite{brey}
and that disorder, not accounted for here, plays an important role
in experimental samples.  The most interesting quantity in our 
view is the internal differential capacitance, which shows singular behavior at 
the phase boundary.  Results are plotted here for 
$(\Delta_t,\Delta_z) = (0,0.02)$, (0.02,0.02), and (0.1,0.02).
These choices are motivated by the fact that it is difficult
to fabricate samples with values of the Zeeman coupling in 
interaction units $e^2/\varepsilon\ell$ 
which differ from 0.02 by more than
a factor of two at $\nu=2$, while it is relatively easy 
to vary $\Delta_t$ from
a large value to one which is immeasurably small, by changing
the Aluminum content in the barrier separating the quantum wells.

Numerical calculations for $\Delta_t=0$ and $\Delta_z=0.02$, presented
in Fig.~\ref{numerical}(a), confirm analytical results of Section V:
the order parameter $O_{zz}=0$ while $O_{xz}$ is non-zero for $\Delta_V
>  \Delta_z+2F_- = 0.6176$ and $\Delta_V < \Delta_z+4H-2F_- = 1.4024$;
the quasiparticle excitation gap $E_{gap}$ is constant in the ordered 
region; the charge is equally distributed between layers in the 
low $\Delta_V$ disordered region, all electrons are in the top layer
in the high $\Delta_V$ disordered region and the charge is transferred
between wells linearly with $\Delta_V$ in the ordered region. In 
Fig.~\ref{numerical}(b) we show numerical results for 
$\Delta_t=\Delta_z=0.02$. Both low and high $\Delta_V$ disordered regions
are still present for this choice of external fields. The ordered 
region with both $O_{xz}$ and $O_{zz}$ non-zero is slightly shifted to
lower bias potentials and the step in $C_{int}$ is reduced at the right
phase boundary. Fig.~\ref{numerical}(c) corresponds to large enough
tunneling energy ($\Delta_t=0.1$) so that the $\Delta_V=0$ point falls
into the ordered region of  the zero-bias plane of the phase diagram
(see Fig.~\ref{dtdz}). Hence, only the high $\Delta_V$ phase boundary
is present and is shifted further to lower $\Delta_V$, qualitatively
consistent with
our previous analysis of the $\Delta_z = 0$ phase plane.

\section{Discussion}

The variational ansatz (\ref{generalwf}) on which we have based
the analytic portion our analysis carries only 3 parameters.
One might wonder if more general wavefunctions with more parameters
could unearth lower
energy states. Let us consider how many parameters at most can be
involved in the most general translationally invariant LLL state
of bilayer electrons with spin, at a  filling $\nu = 2 $. Such a state
will correspond, at each spatial point or LLL orbital, to some
2-fermion state where each fermion is described by a 4-spinor.
More generally, consider N fermions, each a d-spinor.  There are
$d! / (N! (d-N)!) $ combinations for placing N  fermions in
d states. Each such combination corresponds to an independent
N-fermion state. These form a basis and a general state will be
a linear combination of these with complex coefficients . Since
each complex coefficient carries two real parameters the total
number of parameters in the most general state will be
$\hat{n} \equiv 2  d! / (N! (d-N)!)  \ - \ 2 $, \  where 2
parameters have been subtracted out which respectively  account
for normalisation and an overall phase. For our problem where d=4
and N=2, $\hat{n} = 10$; much larger than the number
(three) of parameters in our trial spinors.

However not all these $\hat{n}$ independent states can be
written as a single outer product of two 4-spinors, as is
required for the Slater determinant form (\ref{HFstate}) of the
Hartree-Fock trial states.  For example ,  the linear
combination of 2-electron states
\begin{equation} |
T \uparrow,  T \downarrow \rangle \ \  + \ \
| B \uparrow,  B \downarrow \rangle \label{funny} \end{equation}
(where T,B refer to the two layers) cannot
be written as an outer product of two 4-spinors of the form
\begin{equation} \pmatrix{a_1\cr a_2 \cr a_3 \cr a_4\cr}
\qquad   \otimes  \ \ \ \ \ \pmatrix{b_1\cr b_2 \cr b_3 \cr b_4\cr}
\end{equation}
for any values of the constants $a_i \  and  \ b_i$ which would need to
satisfy the conflicting conditions
$ a_1 b_2 = a_3 b_4 = 1$, while $a_1 b_4 = a_3 b_2 = 0$.
Whether such variational states which, for each spatial point, are
\underline{sums} of products of spinors (as for instance in (\ref{funny}))
can be candidates for the ground state of different phases is an
interesting question. ${\it A  priori}$ they are not ruled out by
any basic superselection rules.  However, broken symmetry states
with which we are familiar, for example superconducting and magnetic
states, can be described using an order parameter field 
functional integral language in which a saddle point approximation to
the order parameter functional integral corresponds to a Hartree-Fock
approximation for the broken symmetry ground state.  We assume that
to be the case for double-layer quantum Hall states as
well\cite{joglekar}.  In particular, we believe that the set of 
possible broken symmetries can be identified by looking only at single
Slater determinant states.  The positions of the phase boundaries 
will be altered by quantum and, at finite temperature, also thermal
fluctuations in the order parameter field.  We therefore are 
motivated to identify the number of parameters needed to characterize
all possible Hartree-Fock states.

A product of N orthonormal d-plets will carry
Nd complex parameters to start with. This will be reduced by
$N$ normalization constraints, $N (N-1)/2$ complex orthogonality
constraints, and $N^2$ symmetry constraints since unitary transformations among
the $N$ the single-particle states will not yield a distinct
N-fermion state. This leaves behind $n  \equiv
2dN \ - \ N- 2 N(N-1)/2  \ - N^2  \ =  \ 2 N(d-N)$  real parameters.
Notice that this is just the dimensionality of the coset space
$U(d) / ( U(N) \times U(d-N)) $.  (See Arovas et al \cite{Arovas} who
also do this counting).
For $d=4$ and $N=2$ the number of free
parameters in Hartree-Fock states is $n \ = \ 8$;
for $N=1$, the case relevant to double-layer
systems at $\nu=1$,  $ n = 6$.
For our problem, the number of free parameters is further reduced
by symmetries of the
interaction Hamiltonian which is invariant under arbitrary
(3-parameter) rotations in
spin-space and under (1 more parameter) rotations about the $\hat z$ axis in
pseudospin space.  Of course these symmetries are partially broken by
one-body terms in the Hamiltonian.  However, such breaking simply selects
a known preferred direction for the total spin-polarization (along
the Zeeman coupling direction )
and for the projection of the
total pseudospin polarization onto the $\hat x-\hat y$ plane
(the $\hat x$ direction.)  The end result is that the number of
free parameters is $4$ for $\nu=2$.  We have used 3 parameters based
on the assumption that in the grounds state pseudospin projections
in {\it each} layer will be in the $\hat x  - \hat z$ plane.

\section{Concluding Remark}

The richness of order-disorder physics in double-layer quantum
Hall systems at $\nu=2$ suggests that this should be an 
inviting system for experiment.  To date, only
transport\cite{ezawarecent} and optical\cite{pellegrini} studies,
which probe the state rather indirectly have been completed.
We believe that capacitance studies of double-layer 
systems as a function of an interlayer bias potential will prove
more interesting.  In particular we expect the differential capacitance,
an experimentally accessible quantity which is the analog for these
zero-temperature phase transitions of the specific heat in
thermal phase transitions, to be a very useful probe in 
studying changes in singular dependence on bias potential 
with temperature, Zeeman coupling strength, and 
interlayer tunneling amplitude.

\section{Acknowledgements}

R.R. acknowledges the generous hospitality of 
the Physics Department at Indiana University which led to the 
initiation of this work.
R.R. and A.H.M. acknowledge the hospitality of 
the Institute for Theoretical Physics (ITP) at UC Santa
Barbara while this work was in progress.
Research at ITP was supported by NSF grant NO. PHY 94-07194.
The work was also supported by the National Science Foundation under
Grants DMR-9623511, DMR-9714055, and INT-9602140, by the
Ministry of Education of the Czech Republic under Grant No. ME-104, and
by the Grant Agency of the Czech Republic under Grant No. 202/98/0085.

\begin{figure}
\caption{Schematic of the conduction band edge profile
and one-particle energy levels
in a biased double quantum well structure.  The four one-particle
states are products of orbitals split due to both interlayer tunneling
and a bias potential, and spinors split by the Zeeman coupling.
Interactions become important and broken symmetry states occur 
at $\nu=2$ when the second and third levels, and sometimes also the
first and fourth levels, are close to degeneracy} 
\label{doublewell}
\end{figure}

\begin{figure}
\caption{$\Delta_z=0$ plane of the phase diagram.  The ground
state has SU(2) broken symmetry throughout the 
shaded region and U(1)$\otimes$SU(2) broken symmetry along the darkened
portion of the $\Delta_z=0 - \Delta_t=0$ line.
The dashed-line arrows labeled $(i,\sigma)$ illustrate schematically
the pseudospin of spin-$\sigma$ projections of the 
spinor $z^i$ at a given point in the phase
diagram.  An upward arrow represents an orbital localized in the top layer,
downward arrow an orbital in the bottom layer, while a finite tilt from
the vertical direction indicates interlayer phase coherence.
The solid-line arrows show schematically the orientation of the
spin polarization in top (upper box) and bottom (lower box) quantum
wells. The length of the arrow represents
the magnitude of the spin polarization in that layer while 
its thickness represents the fraction of the charge density.}
\label{dtdv}
\end{figure}

\begin{figure}
\caption{$\Delta_V = 0$ plane of the phase diagram.  The ground state
has a U(1) broken spin symmetry in the shaded region.
The dashed-lines and solid-lines summarize respectively the 
spin and pseudospin structures of the ground state as in Fig.~2.}
\label{dtdz}
\end{figure}
\begin{figure}
\caption{$\Delta_t=0$ plane of the phase diagram.
The dashed-lines and solid-lines summarize respectively the 
spin and pseudospin structures of the ground state as in Fig.~2.
Spontaneous interlayer phase
coherence is developed in the shaded region.}
\label{dvdz}
\end{figure}

\begin{figure}
\caption{3D phase diagram for bilayer system at $\nu=2$. Ordered states
occur in the volume enclosed by the shaded and transparent surfaces.}
\label{3d}
\end{figure}

\begin{figure}
\caption{Order parameters, quasiparticle excitation gap, charge distribution,
and internal capacitance as a function of bias potential for $\Delta_t=0$
and $\Delta_z=0.02$ (a), $\Delta_t=\Delta_z=0.02$ (b), and $\Delta_t=0.1$
and $\Delta_z=0.02$ (c).}
\label{numerical}
\end{figure}

\begin{thebibliography}{10}
\narrowtext

\bibitem{ahmintro}
A.H. MacDonald,  in {\em Proceedings of the 1994 Les Houches Summer
School on Mesoscopic Quantum Physics}, edited by E. Akkermans
{\it et~al.} (Elsevier Science, Amsterdam, 1995), pp.\ 659--720.

\bibitem{qhferro} S.L. Sondhi, A. Karlhede, S.A. Kivelson, and E.H.
Rezayi, Phys. Rev. B {\bf 47}, 16419 (1993);
A.H. MacDonald, H.A. Fertig, and L. Brey, Phys. Rev. Lett.
{\bf 76}, 2153 (1996).

\bibitem{zeeman} In the semiconductor systems in which the
quantum Hall effect occurs, the Zeeman coupling of a magnetic field to the
spin-degree of freedom is approximately 60 times weaker than
coupling to the electronic orbital degrees of freedom.
Zeeman coupling is weak compared to electron-electron interactions
in the quantum Hall regime.  The coupling is nevertheless not
zero, unless carefully tuned by adjusting pressure or quantum
well width, and ultimately determines the direction of
the ordered spin-moments at $\nu=1$.

\bibitem{dlbrokensym} H.A. Fertig, Phys. Rev. B {\bf 40}, 1987 (1989);
A.H. MacDonald, P.M. Platzman, and G.S. Boebinger,
Phys. Rev. Lett. {\bf 65}, 775 (1990);
X.G. Wen and A. Zee, Phys. Rev. Lett. {\bf 69}, 1811 (1992);
Z.F. Ezawa and A. Iwazaki, Int. J. Mod. Phys. B {\bf 6}, 3205 (1992);
S.M. Girvin and A.H. MacDonald, ``Multicomponent Quantum Hall
Systems: The Sum of  Their Parts and More", in {\it Perspectives in
Quantum Hall Effects}{Wiley, New York, 1997}, edited by S. Das Sarma
and A. Pinczuk and work cited therein.

\bibitem{zhengdassarma} L. Zheng, R.J. Radtke, and S. Das Sarma,
Phys. Rev. Lett. {\bf 78}, 2453 (1997); S. Das Sarma, S. Sachdev and L. Zheng,
Phys. Rev. Lett. {\bf 79}, 917 (1997); Phys. Rev. B {\bf 58}, 4672 (1998).

\bibitem{pellegrini} V. Pellegrini, A. Pinczuk, B.S. Dennis, A.S. Plaut,
L.N. Pfeiffer, and K.W. West, Phys. Rev. Lett. {\bf 78}, 310 (1997);
Science {\bf 281}, 799 (1998);

\bibitem{joglekar} Y. Joglekar and A.H. MacDonald,
in preparation (1999).

\bibitem{swierkowski} L.\'Swierkowski and A.H. MacDonald, Phys. Rev. B {\bf 55},
R16017, (1997).

\bibitem{harfok} A.H. MacDonald, Phys. Rev. B {\bf 30}, 4392 (1984); 
A.H. MacDonald, H.C. Oji, and K.L. Liu,
Phys. Rev B {\bf 34}, 2681 (1986).

\bibitem{remark} Properties change qualitatively when the sense of
this inequality is altered.  This can happen, for example, when
the Landau level index is different in the two quantum wells:
T. Jungwirth, S.P. Shukla, L. Smr\v{c}ka, M.. Shayegan, and A.H. MacDonald,
Phys. Rev. Lett. {\bf 81}, 2328 (1998).

\bibitem{ezawarecent} A. Sawada, A.F. Ezawa, H. Ohno, Y. Horikoshi,
Y. Ohno, S. Kishimoto, F. Matsukura, Phys. Rev. Lett. {\bf 80}, 4534 (1998).

\bibitem{capexp} J.P. Eisenstein, L.N. Pfeiffer, and K.W. West,
Phys. Rev. B {\bf 50}, 1760 (1994); Phys. Rev. Lett. {\bf 68}, 674 (1992);
T. Jungwirth and A.H. MacDonald, Phys. Rev. B {\bf 53}, 9943 (1996).

\bibitem{doublelayerrefs} K. Moon, H. Mori, K. Yang, S.M. Girvin,
and A.H. MacDonald, Phys. Rev. B {\bf 51}, 5138 (1995);
K. Yang, K. Moon, L. Belkhir, H. Mori, S.M. Girvin, A.H. MacDonald,
L. Zheng, and D. Yoshioka, {\em ibid} {\bf 54}, 11644 (1996).

\bibitem{demler} L. Brey, E. Demler, and S. Das Sarma,
preprint, cond-mat/9901296, have completed a careful comparison between
Hartree-Fock theory predictions for the energy gap and 
recent experiments.

\bibitem{brey} B. Paredes, C. Tejedor, L. Brey, and L. Martin-Moreno,
preprint, cond-mat/9902197.

\bibitem{Arovas} D.P. Arovas, A. Karlhede, and D. Lillihook,
preprint, cond-mat/9811097.

\end{thebibliography}
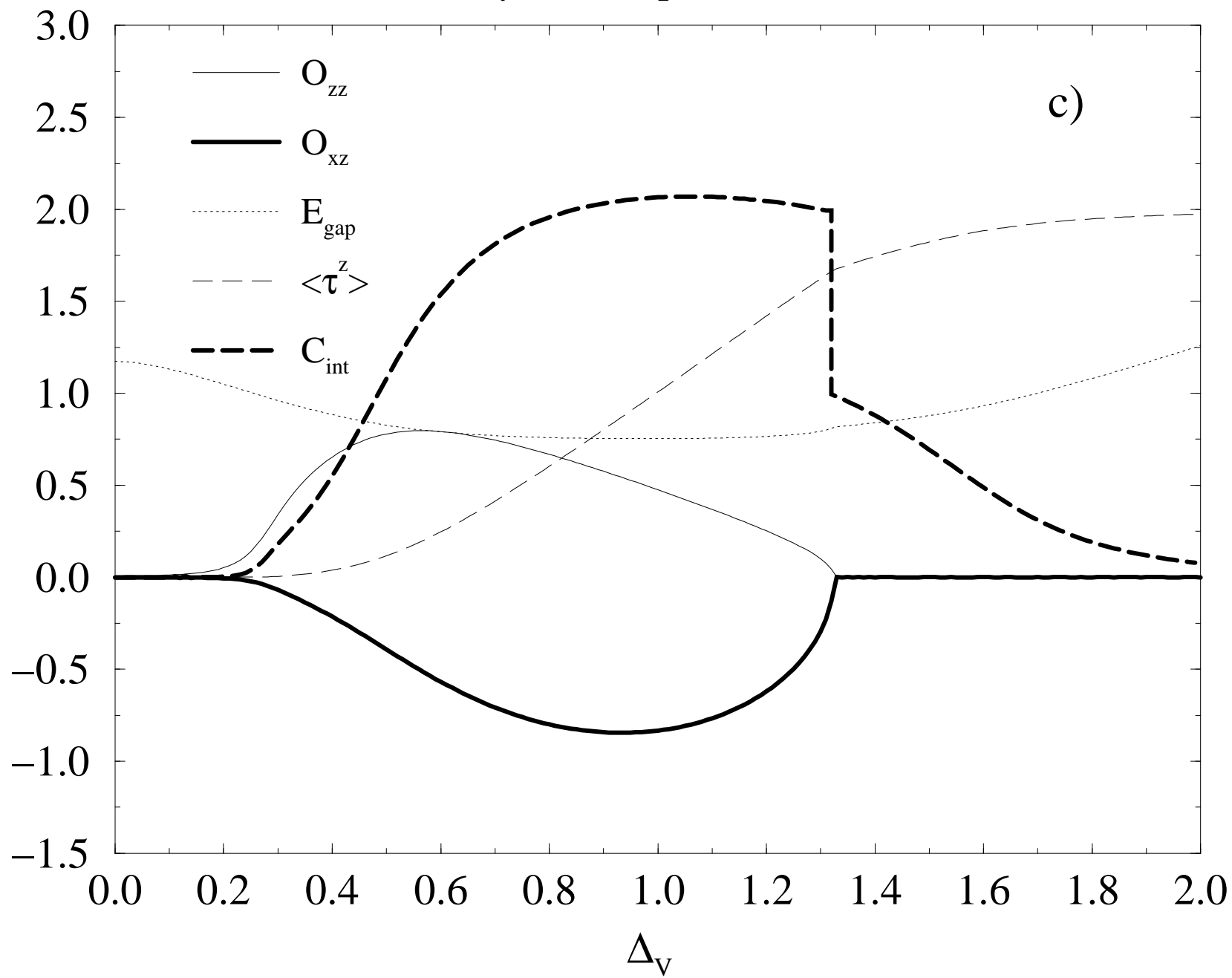
\end{document}